\documentclass[12pt,letterpaper]{article}
\usepackage{amsmath,amssymb,array,calc,rotating,epsfig,psfrag,amscd, cite}

\numberwithin{equation}{section}

\makeatletter
\renewcommand\section{\@startsection {section}{1}{\z@}%
                                   {-3.5ex \@plus -1ex \@minus -.2ex}
                                   {2.3ex \@plus.2ex}%
                                   {\normalfont\large\bfseries}}
\renewcommand\subsection{\@startsection{subsection}{2}{\z@}%
                                     {-3.25ex\@plus -1ex \@minus -.2ex}%
                                     {1.5ex \@plus .2ex}%
                                     {\normalfont\bfseries}}
\makeatother

\let\S=\Sigma

\newcommand{\bea}{\begin{eqnarray}}
\newcommand{\eea}{\end{eqnarray}}
\newcommand{\be}{\begin{equation}}
\newcommand{\ee}{\end{equation}}

\newcommand{\p}{\partial}

\newcommand{\C}[1]{$(\ref{#1})$}

\def\IZ{\relax\ifmmode\mathchoice
{\hbox{\cmss Z\kern-.4em Z}}{\hbox{\cmss Z\kern-.4em Z}}
{\lower.9pt\hbox{\cmsss Z\kern-.4em Z}} {\lower1.2pt\hbox{\cmsss
Z\kern-.4em Z}}\else{\cmss Z\kern-.4em Z}\fi}
\def\IR{\relax{\rm I\kern-.18em R}}

\def\one{{\hbox{ 1\kern-.8mm l}}}

\newlength{\bredde}
\def\slash#1{\settowidth{\bredde}{$#1$}\ifmmode\,\raisebox{.15ex}{/}
\hspace*{-\bredde} #1\else$\,\raisebox{.15ex}{/}\hspace*{-\bredde}
#1$\fi}

\newsavebox{\zzzbar}
\sbox{\zzzbar}
  {\setlength{\unitlength}{0.9em}
  \begin{picture}(0.6,0.7)
  \thinlines
  \put(0,0){\line(1,0){0.6}}
  \put(0,0.75){\line(1,0){0.575}}
  \multiput(0,0)(0.0125,0.025){30}{\rule{0.3pt}{0.3pt}}
  \multiput(0.2,0)(0.0125,0.025){30}{\rule{0.3pt}{0.3pt}}
  \put(0,0.75){\line(0,-1){0.15}}
  \put(0.015,0.75){\line(0,-1){0.1}}
  \put(0.03,0.75){\line(0,-1){0.075}}
  \put(0.045,0.75){\line(0,-1){0.05}}
  \put(0.05,0.75){\line(0,-1){0.025}}
  \put(0.6,0){\line(0,1){0.15}}
  \put(0.585,0){\line(0,1){0.1}}
  \put(0.57,0){\line(0,1){0.075}}
  \put(0.555,0){\line(0,1){0.05}}
  \put(0.55,0){\line(0,1){0.025}}
  \end{picture}}

\newcommand{\ena}{\end{eqnarray}}
\newcommand{\beqa}{\begin{eqnarray}}
\newcommand{\eeqa}{\end{eqnarray}}

\def\S{\Sigma}

\begin{document}
\begin{titlepage}

\begin{center}



\vskip 2 cm
{\Large \bf Algebraic identities between families of (elliptic) modular graphs}\\
\vskip 1.25 cm { Anirban Basu\footnote{email address:
    anirbanbasu@hri.res.in} } \\
{\vskip 0.5cm  Harish--Chandra Research Institute, A CI of Homi Bhabha National
Institute, \\ Chhatnag Road, Jhusi, Prayagraj 211019, India}

\end{center}

\vskip 2 cm

\begin{abstract}
\baselineskip=18pt

Consider an algebraic identity between elliptic modular graphs where several vertices are at fixed locations (and hence unintegrated) while the others are integrated over the toroidal worldsheet. At any unintegrated vertex, we can glue an arbitrary expression involving elliptic modular graphs which has the same unintegrated vertex. Integrating over that vertex, we obtain new algebraic identities between elliptic modular graphs. Hence this elementary process of convoluting the original ``seed'' identity with other graphs yields infinite number of new identities. We consider various seed identities in which two of the vertices are unintegrated. Convoluting them with families of elliptic modular graphs, we obtain new identities. Each identity is parametrized by an arbitrary number of links in the graphs as well as the positions of unintegrated vertices. On identifying the unintegrated vertices, this leads to an algebraic identity involving modular graphs where all the vertices are integrated over the worldsheet.

\end{abstract}

\end{titlepage}


\section{Introduction}

Modular graph functions~\cite{DHoker:2015gmr,DHoker:2015wxz} arise in the analysis of the low momentum expansion of one loop amplitudes in string theory. These $SL(2,\mathbb{Z})$ invariant graphs have links given by the scalar Green function on the toroidal worldsheet, where one of the vertices is at a fixed location and the others are integrated over the torus. These can be generalized to elliptic modular graph functions~\cite{DHoker:2015wxz,DHoker:2017pvk,DHoker:2018mys} where, in the simplest case, two of the vertices are at fixed locations on the torus. In fact, they arise in the asymptotic expansion of two loop string amplitudes around the non--separating node on the moduli space of genus two Riemann surfaces. Thus while modular graphs are functions of $\tau$, the complex structure of the torus, these elliptic modular graphs are functions of $\tau$ and $v$, where the two unintegrated vertices are at locations $v$ and $0$. They are invariant under the $SL(2,\mathbb{Z})$ transformation
\be \tau \rightarrow \frac{a\tau+b}{c\tau+d}, \quad v \rightarrow \frac{v}{c\tau+d},\ee      
where $a,b,c,d \in \mathbb{Z}$ and $ad-bc=1$.

Now the modular graphs and their elliptic generalizations satisfy eigenvalue equations as well as algebraic identities among themselves which lead to a rich underlying structure (see the reviews~\cite{Berkovits:2022ivl,Dorigoni:2022iem} for various details) demonstrating that not all of them are independent. The algebraic identities satisfied by these graphs have been obtained for graphs with a fixed number of links (see~\cite{DHoker:2015gmr,DHoker:2016mwo,Basu:2016kli,DHoker:2016quv,Broedel:2018izr,DHoker:2019blr,Gerken:2020aju} for such relations). Of course, it would be interesting to obtain algebraic identities between graphs where the number of links is arbitrary, and in the case of elliptic graphs, where the number of vertices fixed on the worldsheet (and hence unintegrated) are arbitrary. This is important in order to construct a basis of independent graphs. In this short note, we report some results along this direction using elementary means.

To start with, we see that an identity involving only modular graphs cannot lead to any more identities as all the vertices are integrated\footnote{Using translational invariance, one of the vertices can be at a fixed location, but the same conclusion is true.}. However, things are very different when we consider an identity involving elliptic modular graphs. Let us denote such an identity by ($n \geq 2$)
\be \label{seed}F(v_1, \ldots, v_n)=0,\ee 
where $v_i$ ($i=1, \ldots, n$) are the positions of the unintegrated vertices\footnote{One of them can be at the origin without any loss of generality.} and all other labels are implicit. Let $H(v_j)$ ($1\leq j \leq n$) be any expression involving elliptic modular graphs where all other labels are implicit. Then gluing this expression to the ``seed'' identity at the vertex $v_j$ and integrating it over the worldsheet leads to the new identity
\be \label{new}\int_\S \frac{d^2 v_j}{{\rm Im}\tau} F(v_1, \ldots, v_n) H(v_j) =0.\ee  
In \C{new} and elsewhere, a vertex is integrated with the $SL(2,\mathbb{Z})$ invariant measure
\be \label{m}\frac{d^2z}{{\rm Im}\tau}\ee 
over the toroidal worldsheet $\S$, where $z$, the coordinate on the torus, is given by
\be -\frac{1}{2}\leq {\rm Re} z \leq \frac{1}{2}, \quad 0 \leq {\rm Im}z \leq {\rm Im}\tau.\ee
The integration measure involves $d^2 z = d{\rm Re}z d{\rm Im}z$. Since $H(v_j)$ is arbitrary, this elementary manipulation yields an infinite number of new algebraic identities between the elliptic modular graphs. Thus starting with a seed identity, we obtain new identities by convoluting with other graphs. 

Hence it is important to obtain various seed identities in order to proceed. In this paper, we consider three such identities~\cite{Basu:2020pey,DHoker:2020hlp,Hidding:2022vjf,Basu:2022jok} to obtain explicit results. While the first one has graphs with up to four links, the other two have graphs with up to five links. We consider convoluting them with various families of elliptic graphs to obtain an infinite number of new algebraic identities. Each identity involving a family of graphs is parametrized by an arbitrary number of links as well as the positions of the unintegrated vertices. On identifying the unintegrated vertices, this analysis yields several identities between families of modular graphs where all the vertices are integrated over the toroidal worldsheet.            

\section{Algebraic identities between families of elliptic modular graphs}

We first list the basic building blocks of the various graphs, before proceeding to obtain the identities between families of graphs. 

\begin{figure}[ht]
\begin{center}
\[
\mbox{\begin{picture}(340,75)(0,0)
\includegraphics[scale=.7]{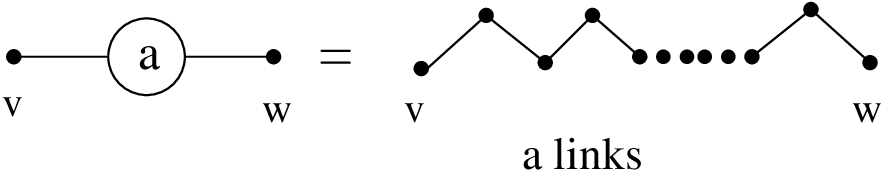}
\end{picture}}
\]
\caption{$G_a(v-w)$, the chain with $a$ links } 
\end{center}
\end{figure}

\subsection{The basic building blocks}

\begin{figure}[ht]
\begin{center}
\[
\mbox{\begin{picture}(400,120)(0,0)
\includegraphics[scale=.65]{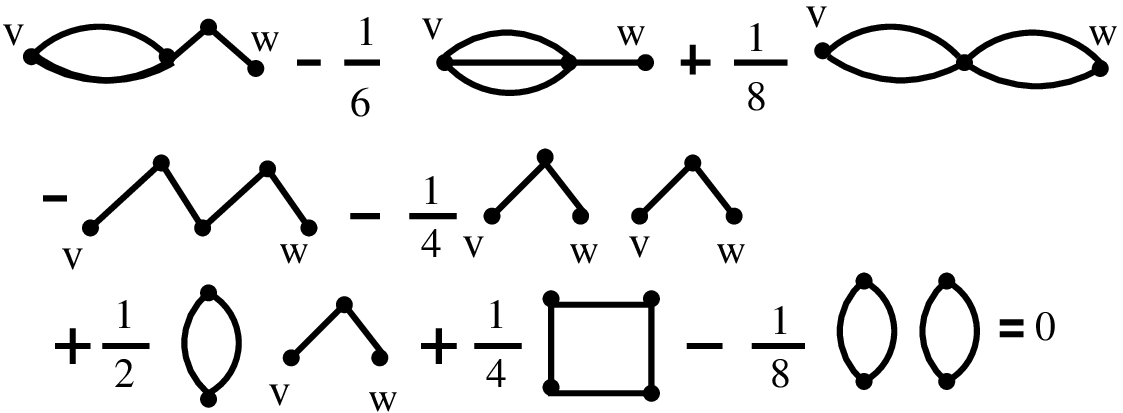}
\end{picture}}
\]
\caption{The first seed identity} 
\end{center}
\end{figure}

\begin{figure}[ht]
\begin{center}
\[
\mbox{\begin{picture}(330,45)(0,0)
\includegraphics[scale=.7]{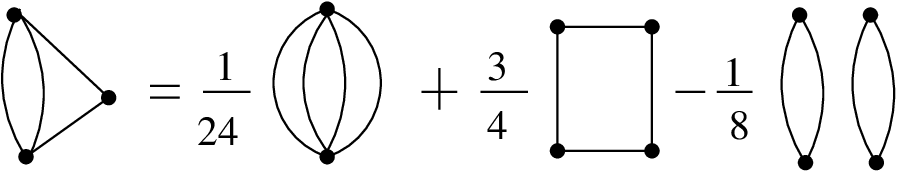}
\end{picture}}
\]
\caption{An identity between modular graphs with up to four links} 
\end{center}
\end{figure}

\begin{figure}[ht]
\begin{center}
\[
\mbox{\begin{picture}(420,85)(0,0)
\includegraphics[scale=.65]{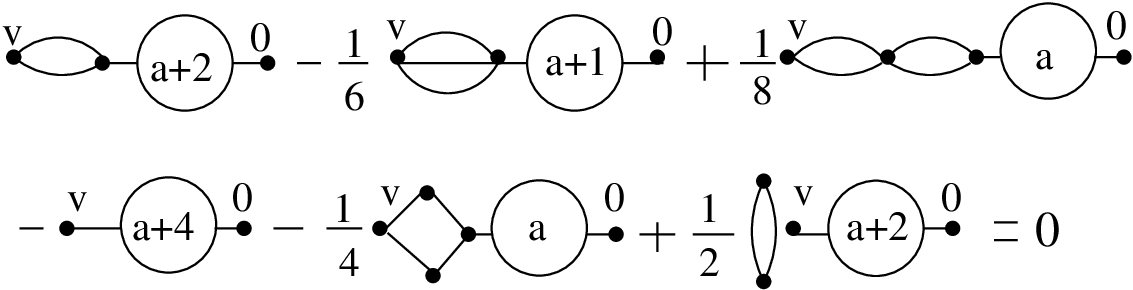}
\end{picture}}
\]
\caption{An identity between elliptic modular graphs with up to $a+4$ links} 
\end{center}
\end{figure}

The basic building blocks of the various graphs are the links which are given by the scalar Green function~\cite{Lerche:1987qk,Green:1999pv}               
\be \label{Green} G(z) = \frac{1}{\pi} \sum_{(m,n) \neq (0,0)} \frac{{\rm Im}\tau}{\vert m\tau+n\vert^2}e^{\pi[\bar{z}(m\tau+n)- z(m\bar\tau+n)]/{\rm Im}\tau},\ee
where we denote $G(z,w;\tau)\equiv G(z-w)$. Thus we have the useful relation
\be \label{v}\int_\S d^2 z G(z-w)=0.\ee
The vertices of the graphs can be either integrated or unintegrated over the toroidal worldsheet. 

In our analysis, it will be very useful to denote the various identities graphically rather than write down their algebraic expressions. In the various graphs, the positions of the unintegrated vertices shall be written down, while the unmarked vertices are the ones that are integrated over. Of course, given a graph one can always revert back to the algebraic expression for it using \C{Green} to express it as a constrained lattice sum.

We shall denote a chain with $a$ links $G_a(v-w)$ ($a\geq 1$) by figure 1 which often arises in our analysis\footnote{Thus $G_1(z) = G(z)$.}. Hence $G_a (0) =E_a$ ($a \geq 2$) which is the non--holomorphic Eisenstein series. In general, identifying all the unintegrated vertices in an elliptic modular graph produces a modular graph.

\subsection{Convoluting seed identities with families of elliptic modular graphs}

We now obtain algebraic identities between families of (elliptic) modular graph functions along the lines of the discussion above.  
To start with, consider the first seed identity in figure 2 which has graphs with up to four links. The labels $v$ and $w$ of the two unintegrated vertices can be freely interchanged\footnote{This is also true of the second and third seed identities in figures 12 and 16 respectively.}. Integrating over $v$ or $w$, the left hand side  vanishes trivially using \C{v}\footnote{This is often the case in the relations that follow, and we shall only mention the non--trivial relations.}. Identifying them yields the non--trivial identity involving modular graphs with up to four links in figure 3. 

\begin{figure}[ht]
\begin{center}
\[
\mbox{\begin{picture}(370,120)(0,0)
\includegraphics[scale=.75]{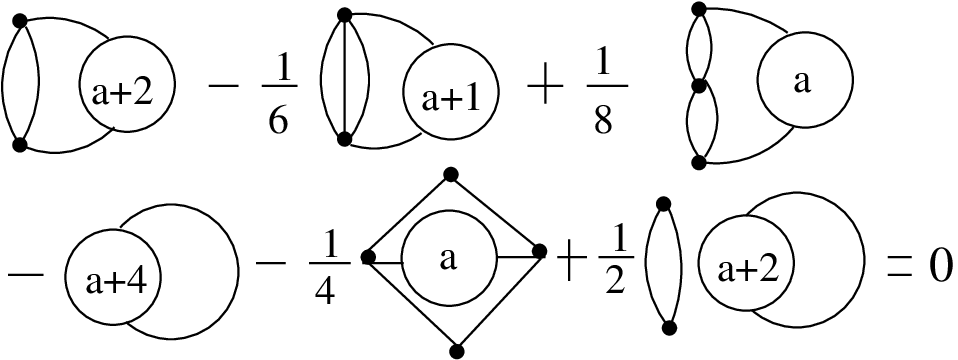}
\end{picture}}
\]
\caption{An identity between modular graphs with up to $a+4$ links} 
\end{center}
\end{figure}

\begin{figure}[ht]
\begin{center}
\[
\mbox{\begin{picture}(180,65)(0,0)
\includegraphics[scale=.75]{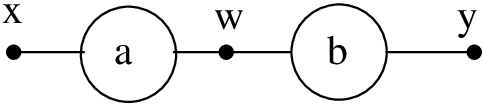}
\end{picture}}
\]
\caption{An elliptic modular graph with $a+b$ links} 
\end{center}
\end{figure}

Now as discussed earlier, from a given seed identity one can obtain new identities by convoluting it with arbitrary elliptic modular graphs, and we shall simply consider various examples involving families of such graphs.  

\begin{figure}[ht]
\begin{center}
\[
\mbox{\begin{picture}(400,200)(0,0)
\includegraphics[scale=.65]{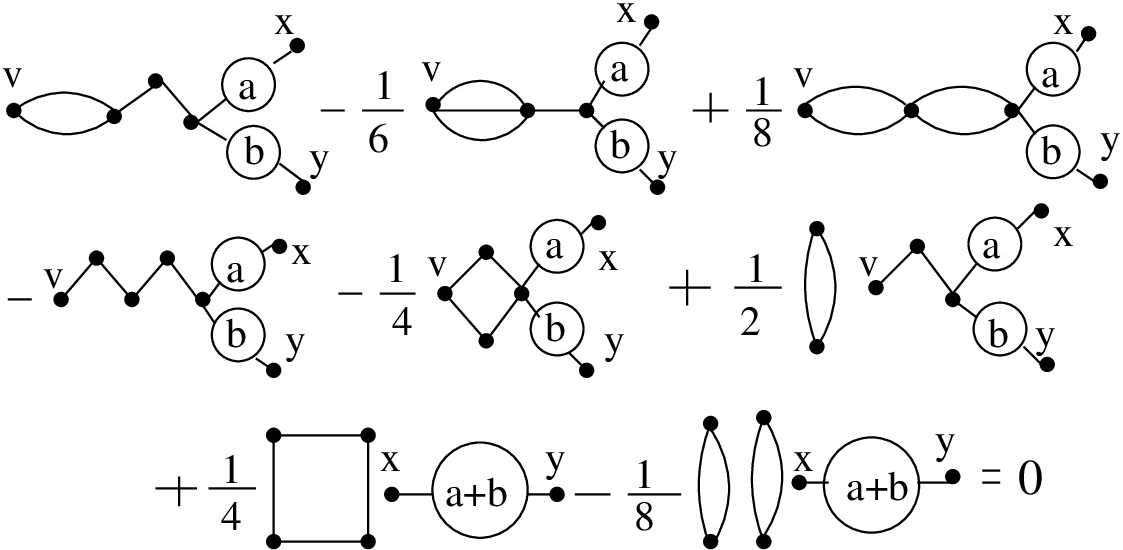}
\end{picture}}
\]
\caption{An identity between elliptic modular graphs with up to $a+b+4$ links} 
\end{center}
\end{figure}

\begin{figure}[ht]
\begin{center}
\[
\mbox{\begin{picture}(410,195)(0,0)
\includegraphics[scale=.75]{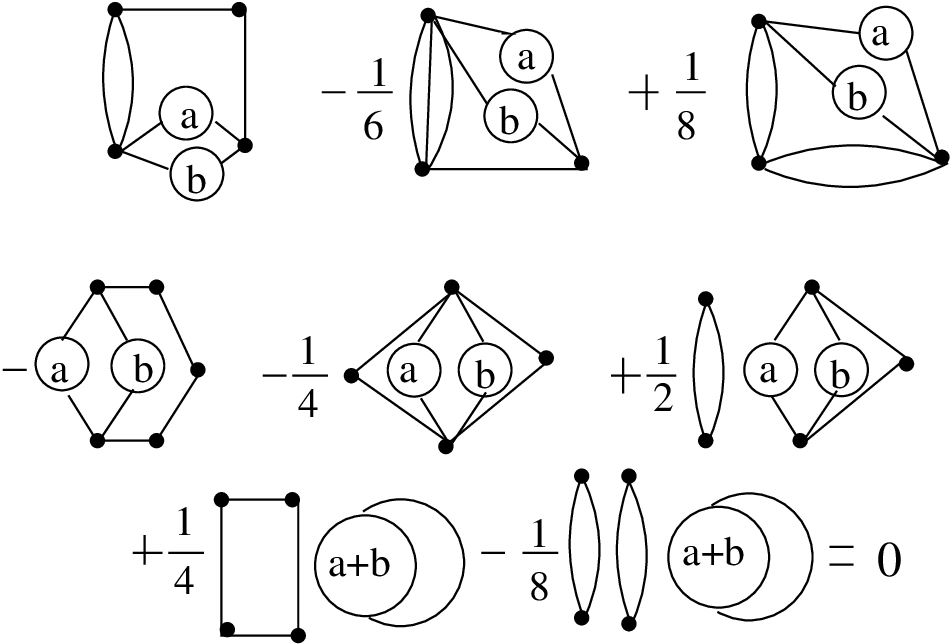}
\end{picture}}
\]
\caption{An identity between modular graphs with up to $a+b+4$ links} 
\end{center}
\end{figure}

\begin{figure}[ht]
\begin{center}
\[
\mbox{\begin{picture}(220,30)(0,10)
\includegraphics[scale=.75]{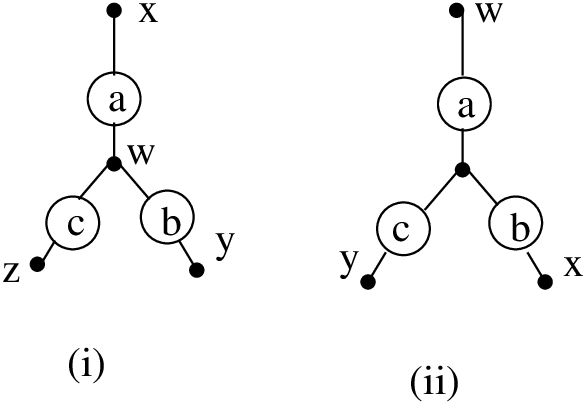}
\end{picture}}
\]
\caption{Elliptic modular graphs with $a+b+c$ links} 
\end{center}
\end{figure}

$\bullet$ To start with, convoluting the first seed identity with $G_a(w)$ ($a \geq 1$) and integrating over the vertex $w$\footnote{Such integrations are always done with the $SL(2,\mathbb{Z})$ invariant measure \C{m}.}, we obtain the identity in figure 4 between elliptic modular graphs with up to $a+4$ links and two unintegrated vertices at $v$ and $0$\footnote{The $a=1$ case reproduces an identity in~\cite{Basu:2022jok}.}.   
Identifying the vertices at $v$ and $0$ leads to the identity between modular graphs with up to $a+4$ links in figure 5. 

$\bullet$ As the next example, we convolute the first seed identity with the elliptic modular graph in figure 6 where $a, b \geq 1$, and integrate over the vertex $w$. This leads to the identity in figure 7  between families of elliptic modular graphs with up to $a+b+4$ links and three unintegrated vertices at $v,x$ and $y$. 
On identifying the vertices $v, x$ and $y$ we obtain the identity between modular graphs with up to $a+b+4$ links in figure 8.  

Thus we have obtained these identities in a very elementary way by convoluting the first seed identity with families of elliptic modular graphs with at most quadratic vertices. While one can convolute with arbitrary families of elliptic modular graphs, we shall simply restrict ourselves to two more examples.     

\begin{figure}[ht]
\begin{center}
\[
\mbox{\begin{picture}(400,300)(0,0)
\includegraphics[scale=.6]{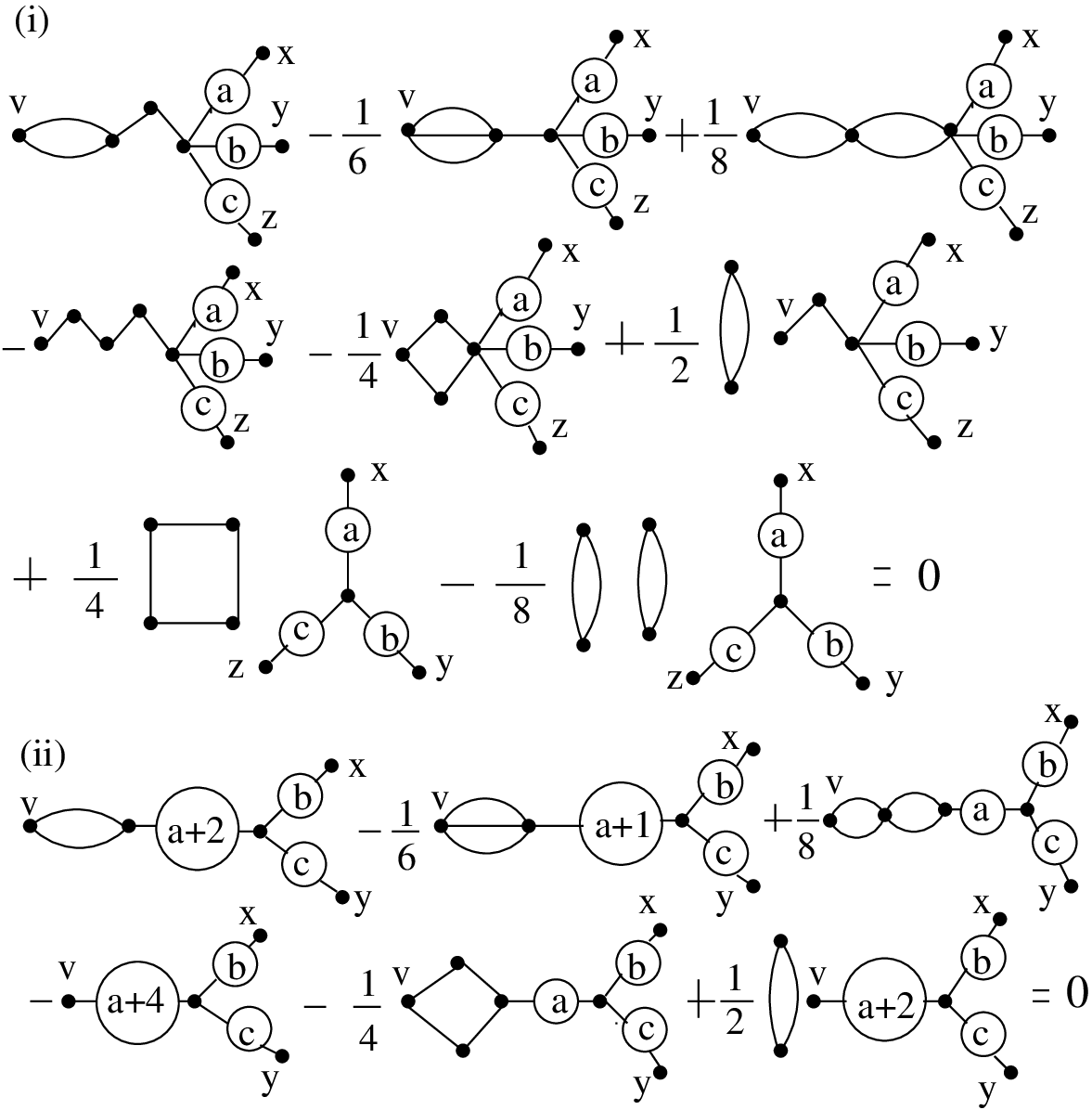}
\end{picture}}
\]
\caption{Identities between elliptic modular graphs with up to $a+b+c+4$ links} 
\end{center}
\end{figure}

$\bullet$ These families of elliptic modular graphs are given in figure 9  where $a,b,c \geq 1$ and we integrate over the vertex $w$ in convoluting with the first seed identity in figure 2. 
Note that instead of convoluting the seed identity with the graph in 9 (ii), we could have alternatively convoluted the identity in figure 4 (by renaming 0 with $w$) with the graph in figure 6 with $a \rightarrow b$, $b \rightarrow c$. These lead to the identities marked (i) and (ii) respectively between elliptic modular graphs in figure 10 with up to $a+b+c+4$ links. While the former has four unintegrated vertices at $v, x, y$ and $z$, the latter has three unintegrated vertices at $v, x$ and $y$.

\clearpage

$\bullet$ On identifying the vertices $v, x, y$ and $z$ in the identity in figure 10 (i), and the vertices $v, x$ and $y$ in the figure in 10 (ii), we obtain identities between families of modular graphs with up to $a+b+c+4$ links, which are given by (i) and (ii) in figure 11 respectively.

\begin{figure}[ht]
\begin{center}
\[
\mbox{\begin{picture}(400,410)(0,0)
\includegraphics[scale=.7]{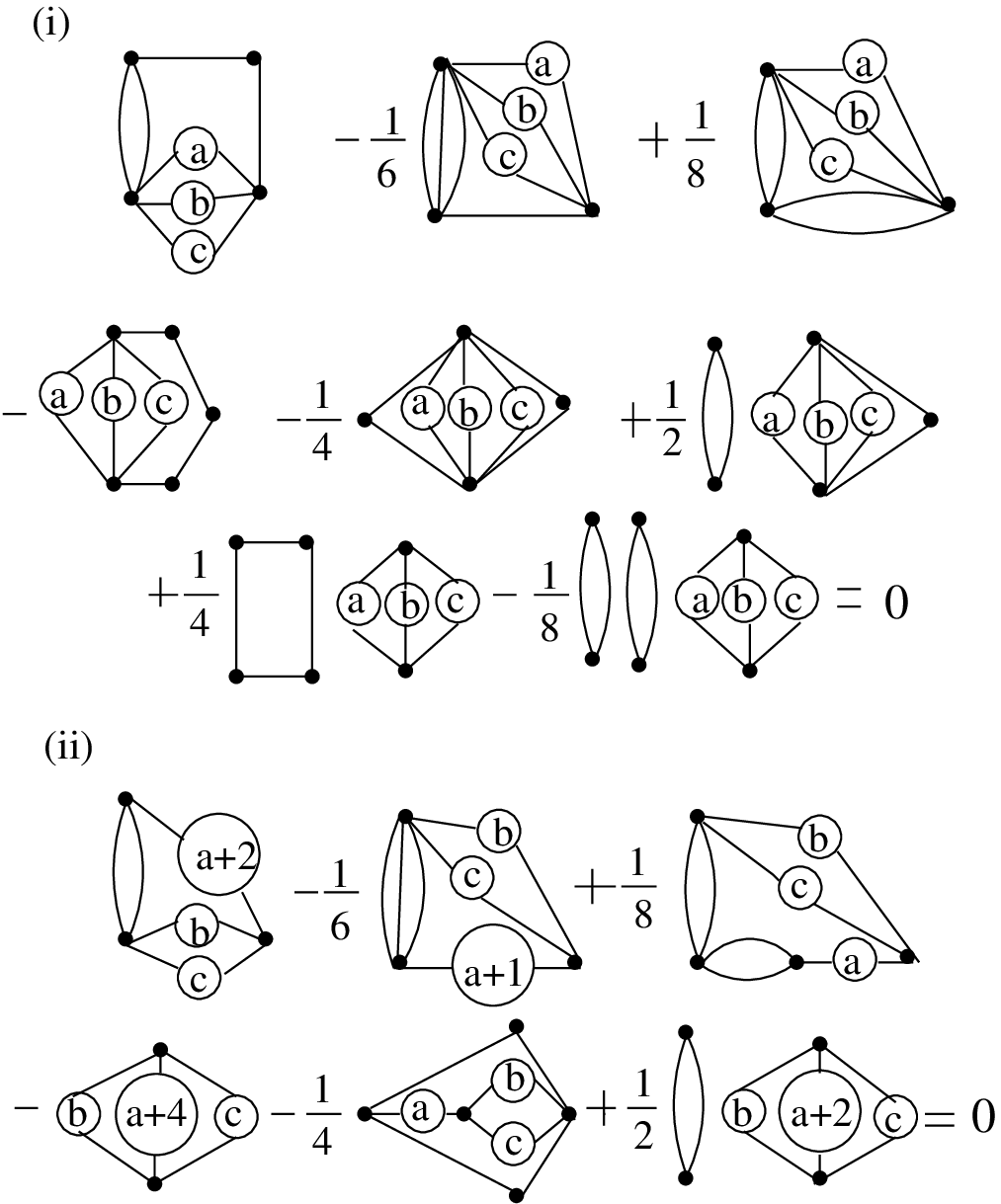}
\end{picture}}
\]
\caption{Identities between modular graphs with up to $a+b+c+4$ links} 
\end{center}
\end{figure}  

\begin{figure}[ht]
\begin{center}
\[
\mbox{\begin{picture}(400,260)(0,0)
\includegraphics[scale=.7]{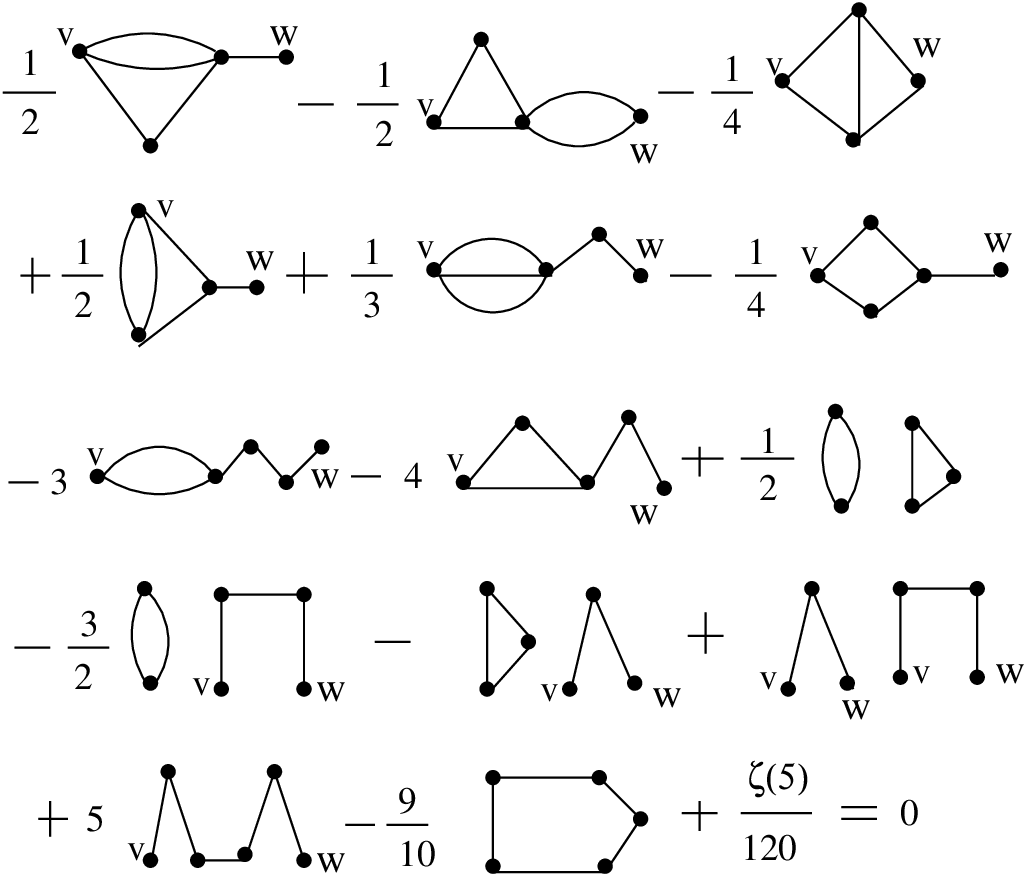}
\end{picture}}
\]
\caption{The second seed identity} 
\end{center}
\end{figure}

\begin{figure}[ht]
\begin{center}
\[
\mbox{\begin{picture}(250,140)(0,0)
\includegraphics[scale=.65]{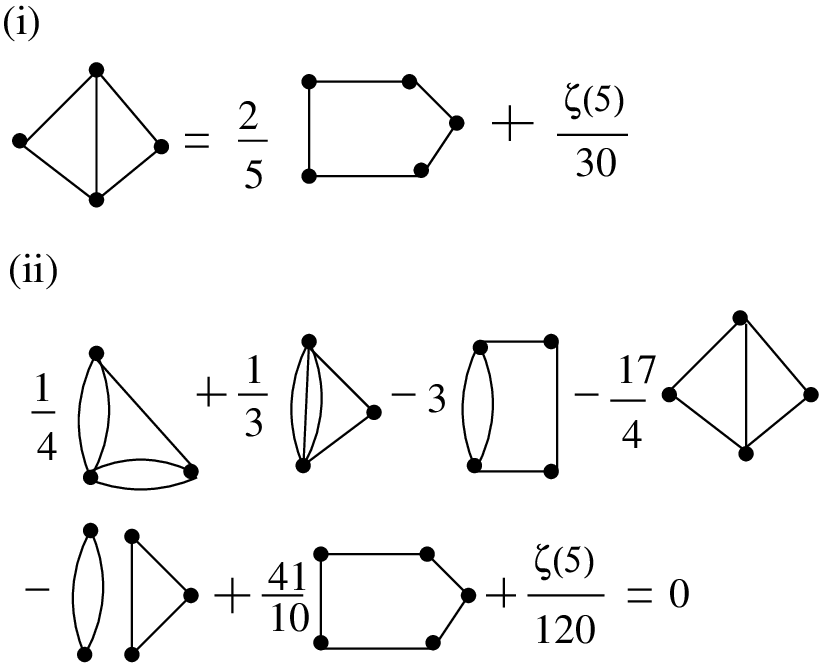}
\end{picture}}
\]
\caption{Identities between modular graphs with up to five links} 
\end{center}
\end{figure}

Hence we see how we can obtain algebraic identities between families of (elliptic) modular graphs using very elementary means. We now consider more examples starting with two more seed identities to obtain more relations. 

\begin{figure}[ht]
\begin{center}
\[
\mbox{\begin{picture}(430,520)(0,0)
\includegraphics[scale=.65]{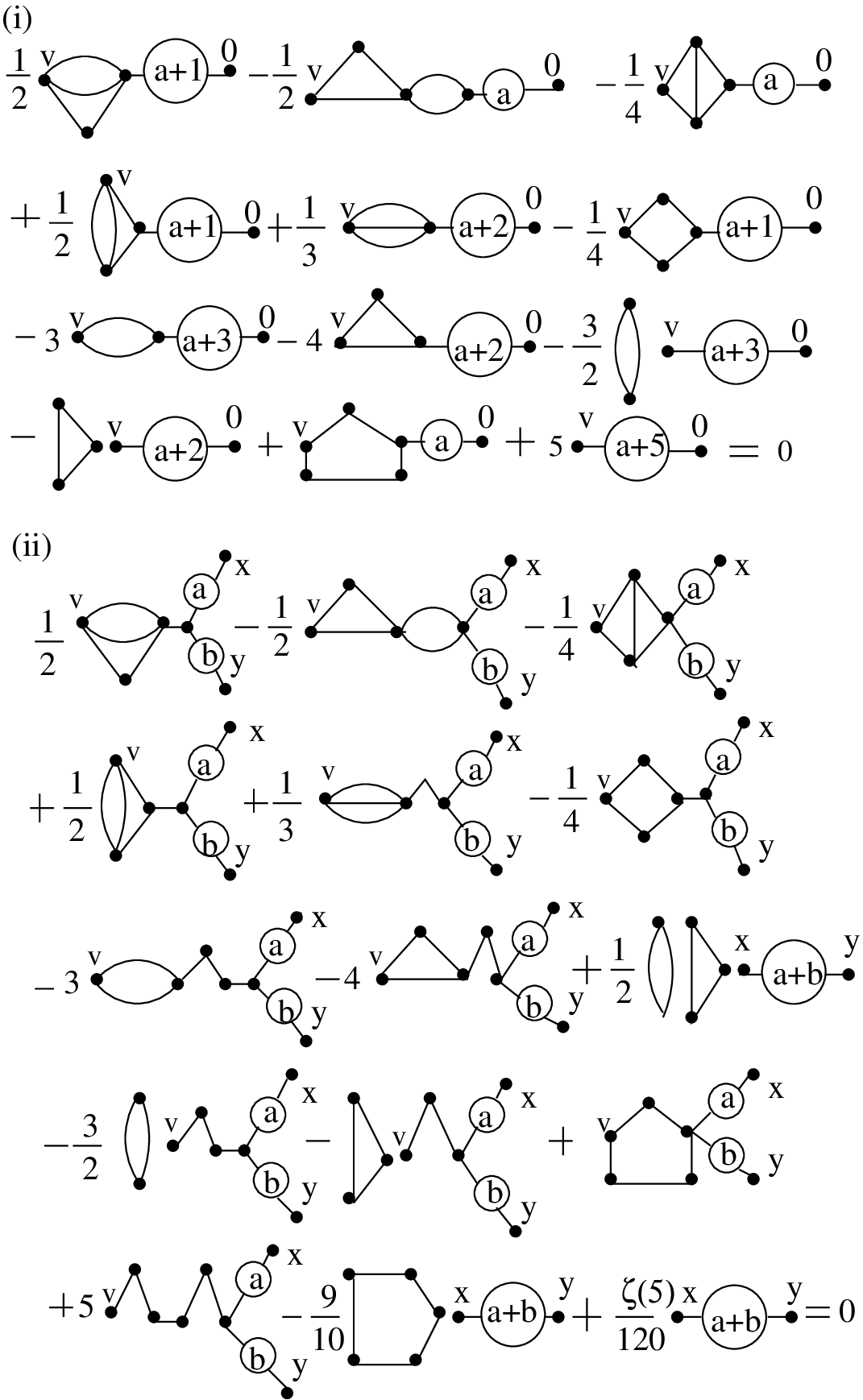}
\end{picture}}
\]
\caption{Identities between elliptic modular graphs with up to (i) $a+5$ links, (ii) $a+b+5$ links} 
\end{center}
\end{figure}

\begin{figure}[ht]
\begin{center}
\[
\mbox{\begin{picture}(460,480)(0,0)
\includegraphics[scale=.6]{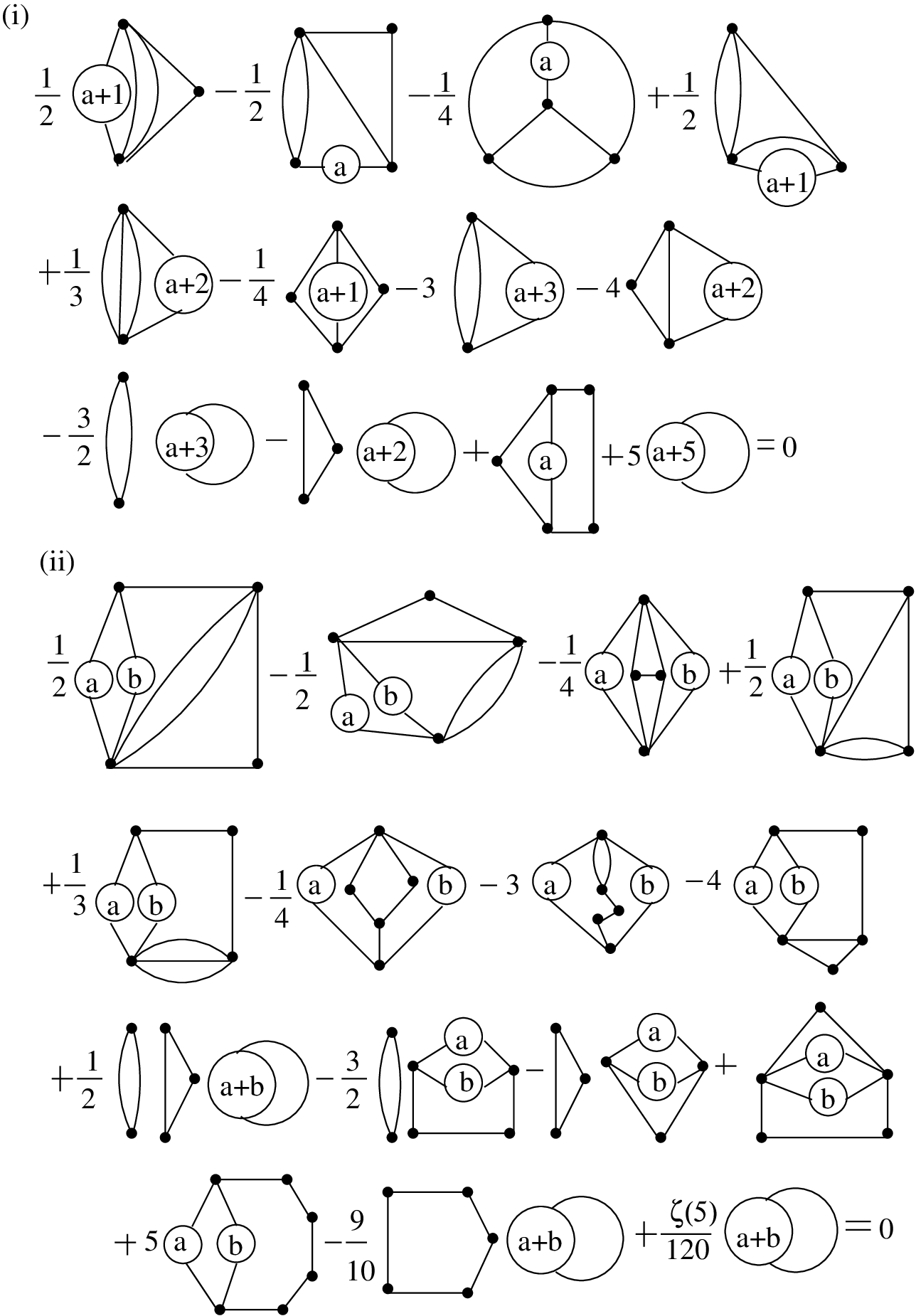}
\end{picture}}
\]
\caption{Identities between modular graphs with up to (i) $a+5$ links, (ii) $a+b+5$ links} 
\end{center}
\end{figure}

\clearpage

Consider the second seed identity in figure 12. While integrating over $v$ or $w$ yields the identity between modular graphs marked (i) in figure 13, identifying the vertices $v$ and $w$ yields the identity marked (ii) in figure 13.   

We now proceed along the lines of the analysis involving the first seed identity and look at some examples. 

$\bullet$ We consider the identities obtained by convoluting the second seed identity with $G_a(w)$ ($a \geq 1$) and integrating over $w$, and also by convoluting by the elliptic modular graph in figure 6 with $a, b \geq 1$. These lead to the identities (i) and (ii) respectively in figure 14 between families of elliptic modular graphs. While the former has graphs with up to $a+5$ links and two unintegrated vertices at $v$ and 0, the later has graphs with up to $a+b+5$ links and three unintegrated vertices at $v, x$ and $y$. 

$\bullet$ Now on identifying the vertices $v$ and $0$ in figure 14 (i), we obtain the identity 15 (i) between families of modular graphs with up to $a+5$ links. Also on identifying the vertices $v, x$ and $y$ in figure 14 (ii), we further obtain the identity 15 (ii) with up to $a+b+5$ links. 
 
\begin{figure}[ht]
\begin{center}
\[
\mbox{\begin{picture}(400,305)(0,0)
\includegraphics[scale=.7]{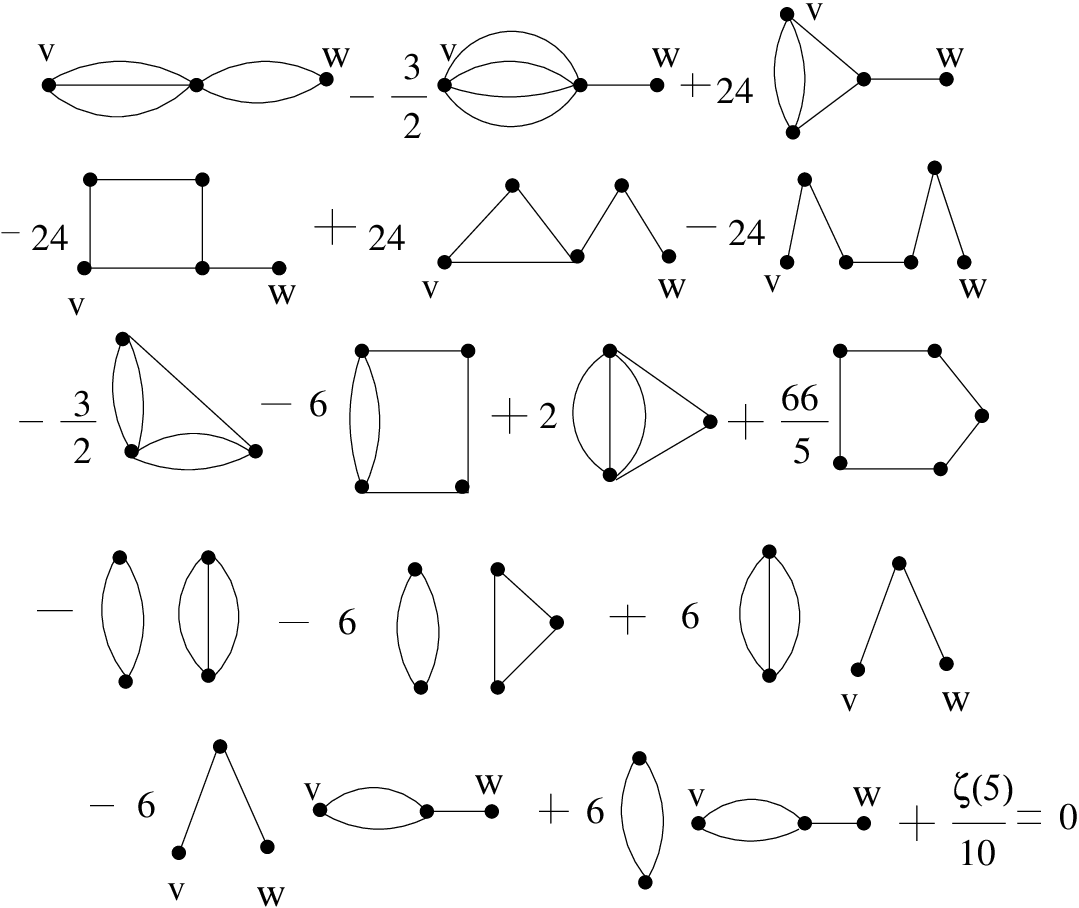}
\end{picture}}
\]
\caption{The third seed identity} 
\end{center}
\end{figure}

\begin{figure}[ht]
\begin{center}
\[
\mbox{\begin{picture}(350,260)(0,0)
\includegraphics[scale=.65]{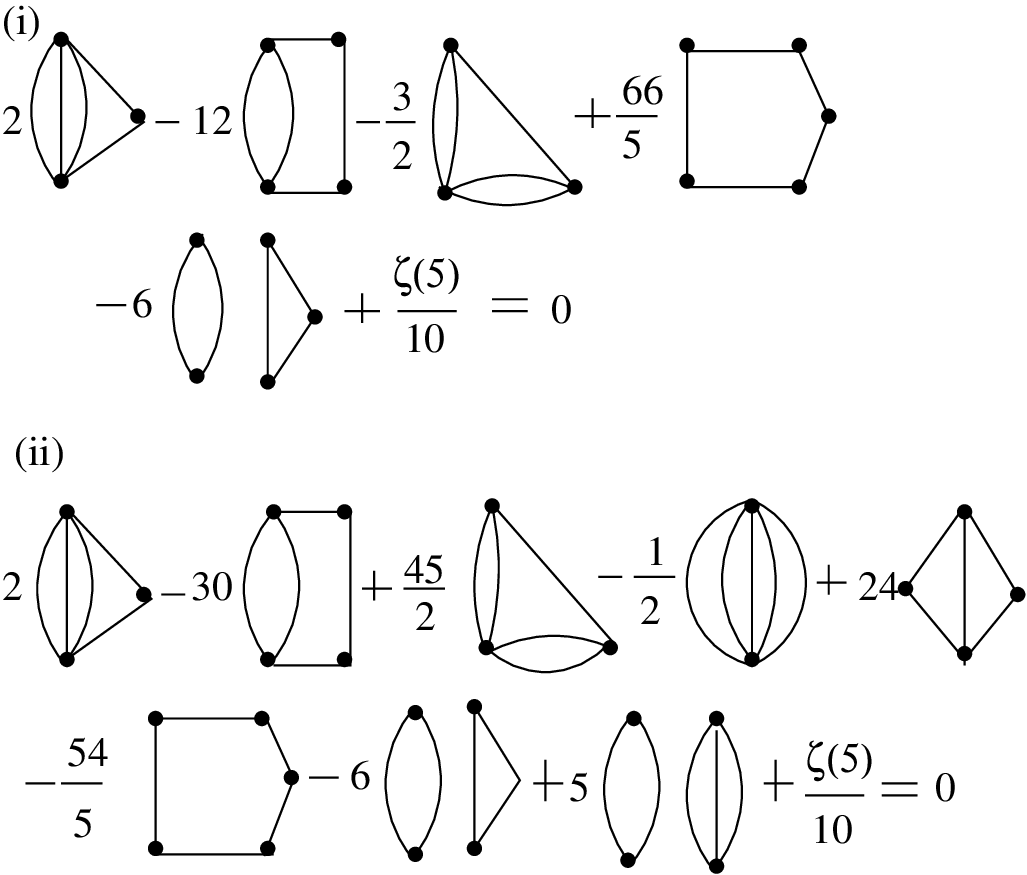}
\end{picture}}
\]
\caption{Identities between modular graphs with up to five links} 
\end{center}
\end{figure}

Finally we consider the third seed identity given in figure 16. While integrating over $v$ or $w$ yields the identity (i) between modular graphs in figure 17, identifying $v$ and $w$ yields the identity marked (ii) in figure 17.   
Note that the graphs in figures 13 and 17 yield four identities between modular graphs with up to five links\footnote{One can check that the identity for $a=1$ in figure 5 does not produce any new relation.}, which can be easily manipulated to reproduce the four known identities between these graphs (see~\cite{DHoker:2015gmr}, for example) providing a non--trivial consistency check.   

$\bullet$ Now proceeding as before, consider the identities obtained by convoluting the third seed identity with $G_a(w)$ ($a \geq 1$) and integrating over $w$, and also by convoluting by the elliptic modular graph in figure 6 with $a, b \geq 1$. These lead to the identities between elliptic modular graphs marked (i) and (ii) respectively in figure 18. Again, while the former has graphs with up to $a+5$ links and two unintegrated vertices at $v$ and 0, the later has graphs with up to $a+b+5$ links and three unintegrated vertices at $v, x$ and $y$. 

$\bullet$
Finally, on identifying the vertices $v$ and $0$ in figure 18 (i), we obtain the 19 (i) between modular graphs with up to $a+5$ links. Also on identifying the vertices $v, x$ and $y$ in figure 18 (ii), we obtain the identity 19 (ii) between modular graphs with up to $a+b+5$ links.  
 
Thus from these examples we see that we have obtained an infinite number of algebraic identities involving families of (elliptic) modular graphs.

While they reproduce all the identities between modular graphs with up to five links they yield a subset of identities involving graphs with up to 
six links and beyond\footnote{Our results agree with those in the literature~\cite{DHoker:2015gmr,DHoker:2016mwo,Basu:2016kli,DHoker:2016quv,Broedel:2018izr,DHoker:2019blr,Gerken:2020aju}.}. 

\begin{figure}[ht]
\begin{center}
\[
\mbox{\begin{picture}(430,450)(0,0)
\includegraphics[scale=.7]{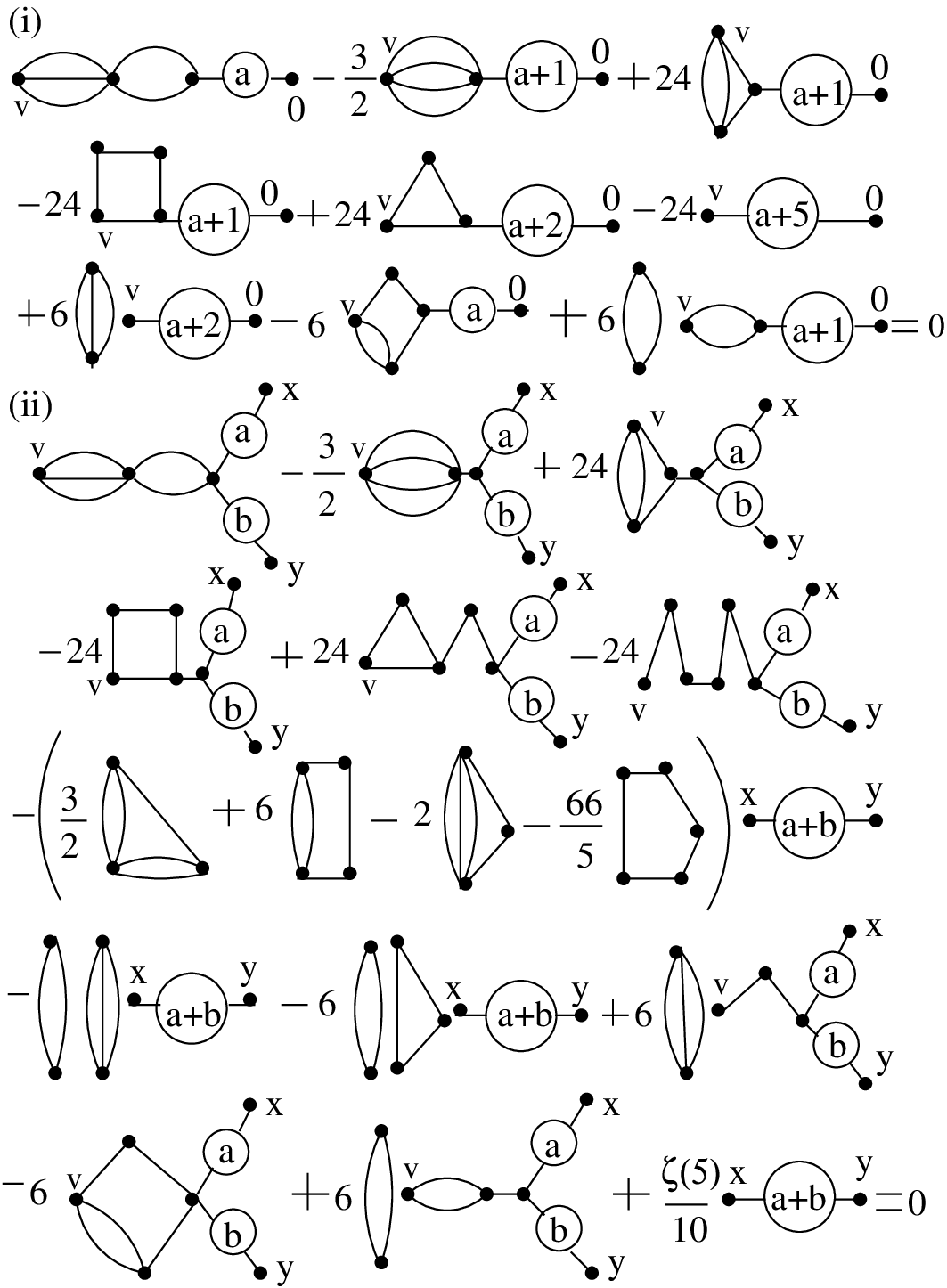}
\end{picture}}
\]
\caption{Identities between elliptic modular graphs with up to (i) $a+5$ links, (ii) $a+b+5$ links} 
\end{center}
\end{figure}

\begin{figure}[ht]
\begin{center}
\[
\mbox{\begin{picture}(410,290)(0,0)
\includegraphics[scale=.55]{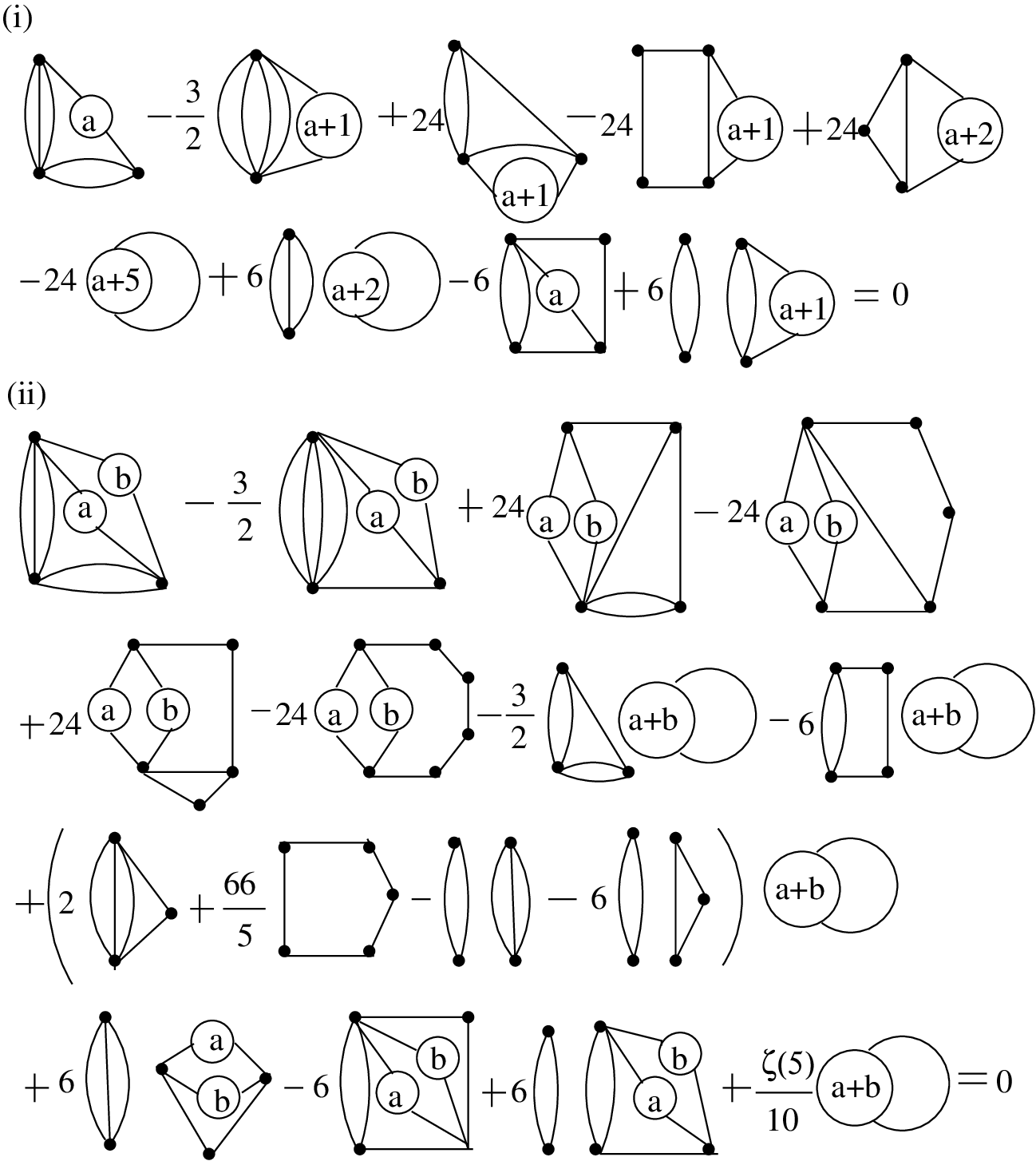}
\end{picture}}
\]
\caption{Identities between  modular graphs with up to (i) $a+5$ links, (ii) $a+b+5$ links} 
\end{center}
\end{figure}

\clearpage

\section{Discussion}

We have outlined a general method to obtain algebraic identities between families of $SL(2,\mathbb{Z})$ invariant (elliptic) modular graph functions starting from seed identities. Though we have illustrated this method with several examples, there are some general conclusions that can be drawn from our analysis as well as possible generalizations, which we very briefly outline.  

Obviously, seed identities play a central role in obtaining the algebraic identities. Thus obtaining seed identities with at least two unintegrated vertices will be a useful exercise. In fact, once these are used to obtain more identities on convoluting with families of elliptic modular graphs as we have done, these can be further used as seed identities to iteratively obtain more algebraic identities involving families of elliptic modular graphs. This easily leads to identities involving graphs with an arbitrary number of links as well as an arbitrary number of unintegrated vertices, which are determined by the choice of the elliptic modular graphs used in convoluting the seed identity. Of course on identifying the unintegrated vertices, we obtain non--trivial identities between families of modular graphs.      

There is a notion of transcendentality which is preserved for every algebraic identity involving families of elliptic modular graphs. This follows from assigning transcendentality one to every link in a graph, while the Riemann zeta function $\zeta (n)$ is assigned the usual transcendentality $n$. For relations involving graphs with sufficiently large number of links, it will be interesting to analyze how generalizations such as multi--zeta values arise as coefficients in the identities.           

While such infinite number of identities between elliptic modular graphs are interesting in their own right, one can ask about their genus two origins. What relations between genus two modular graphs lead to these identities in an asymptotic expansion around the non--degenerating node on their moduli space? In fact, one can possibly think of an iterative structure where we consider relations between families of modular graphs with unintegrated vertices at genus $g$, and then try to obtain their origins in appropriate asymptotic expansions of identities involving modular graphs of genus $g+1$ on their moduli space.          

Finally, the primary strategy used in our analysis is summarized by equations \C{seed} and \C{new}. While we restricted ourselves to $SL(2,\mathbb{Z})$ invariant graphs, the strategy works equally well for modular graph forms~\cite{DHoker:2016mwo}, which are $SL(2,\mathbb{Z})$ covariant graphs. These are graphs where the links are also given by $\p_z G(z,w)$ and $\overline\p_u G(u,v)$, and there are an unequal number of worldsheet $\p$ and $\overline\p$ derivatives sprinkled on the links. Generalizing the issues alluded to above for the case of elliptic modular graph forms should yield a rich and interesting structure. The modular graph functions that arise in our analysis appear in the low momentum expansion of amplitudes in type II string theory, which can be cut open to obtain their elliptic cousins. Likewise it is natural to expect that the analysis involving modular graph forms will involve amplitudes in heterotic string theory.            

\vspace{.5cm}

\noindent {\bf{Acknowledgements:}} I am thankful to the referee for crucial remarks which led to a simple derivation of the result in the earlier version of the draft, and to its generalizations.  


\providecommand{\href}[2]{#2}\begingroup\raggedright\endgroup

\end{document}